# Blockchain Technology Adoption in Food Bank Supply Chains: A Rough DEMATEL-Based Approach

Sara Damavandi[1], Laura Berardi [2], Sina Abbasi[3]

[1]Department of Economic Studies, University of "G. d'Annunzio"  Chieti-Pescara , Italy
sara.damavandi@unich.it
Corresponding Author

[2]Department of Economic Studies, University of "G. d'Annunzio" Chieti-Pescara , Italy
l.berardi@unich.it

Department of Industrial Engineering, Lahijan Branch, Islamic Azad University, Lahijan, Iran
abbasisina170@gmail.com

**Abstract**

Food banks can improve food donation administration, provide real-time inventory tracking, and guarantee compliance with food safety regulations by incorporating blockchain technology. The efficiency, openness, and dependability of food bank supply chains are greatly increased by this integration, leading to more sustainable and successful operations. This study focuses on two primary objectives: identifying key barriers to effective Food bank supply chain (FBSC) operations in blockchain adoption and exploring the interrelationships among these barriers. Barriers were categorized into external and internal frameworks and analyzed using insights from academics and FBs experts. The Decision-Making Trial and Evaluation Laboratory (DEMATEL) methodology was employed to model and quantify the causal relationships among these barriers. DEMATEL's strength lies in its ability to map interdependencies and feedback loops, providing a nuanced understanding of the links between independent and dependent variables in a cause-and-effect network. To address subjectivity and ambiguity in expert opinions during group decision-making, rough theory was integrated with DEMATEL, ensuring a robust approach to handling conflicting perspectives and uncertainty.

**Introduction**

Food banks (FBs) have been established worldwide to address food insecurity while also working to reduce food waste (Dowler & Lambie-Mumford, 2014). They are nonprofit organizations that collect food donations from various sources such as producers, distributors, retailers, consumers, and farmers, and then distribute the food to those in need through a network of community agencies (Tarasuk & Eakin, 2005; Feeding America, 2021). The first FB, St. Mary's Food Bank (SMFB), was founded by John Van Hengel in 1967 in Phoenix, Arizona, USA (Mook et al., 2020). By 2020, FBs affiliated with the European Food Banks Federation (FEBA) had redistributed approximately 860,000 tons of food, helping around 12.8 million people (Akkerman et al., 2023; FEBA, 2021). Food banks are now a well-established component of the charitable food system (CFS), which operates in more than 50 countries and six continents (Global FoodBanking Network, 2022; Riches, 2018). They are essential in helping those who are food insecure.

Together with other European food banks, FEBA plays a significant role in reducing food waste and poverty. The organization relies on corporate donations, public food collection, and mechanisms such as the Fund for European Aid to the Most Deprived (FEAD), which is linked to the European Union's agricultural policies (Capodistrias et al., 2022).

The FAO highlighted the critical role food banks play in international food aid systems in its 2023 report. By allocating excess food that would otherwise go to waste, food banks not only provide immediate relief but also foster sustainability ( FAO, 2023). the absence of comprehensive and explicit guidelines for food safety may lead to the donation of contaminated food or to unsafe donation procedures, endangering the health of recipients who are already at risk.(Mossenson et al., (2024).

likewise, bolstering food banks particularly in poor nations is a key component of the FAO's 2024 strategic plans. In order to better satisfy the nutritional needs of vulnerable populations and support international food security goals, these projects seek to increase the effectiveness of food distribution, improve food safety, and incorporate food banks into larger social protection programs (FAO, 2024).

Recently, food banks have started receiving additional donations from businesses seeking to minimize waste, furthering their role in food recovery and reducing food insecurity (Capodistrias et al., 2022). The operational challenges of food banks include issues with transparency, traceability, and efficient management of food donations. Blockchain technology (BCT) offers a promising solution by providing an immutable, decentralized ledger that ensures transparency and traceability in food donation processes (Tian, 2016; Kamilaris et al., 2019). By utilizing BCT, food banks can track perishable items more effectively, ensuring that they are safely distributed to recipients while reducing waste (Lefebvre, 2021). Furthermore, BCT's ability to create smart contracts can streamline operations by automating agreements among donors, food banks, and recipients, reducing administrative overhead and potential errors (Saberi et al., 2019). Several companies, such as Walmart and Carrefour, have already successfully integrated BCT into their supply chains to improve traceability and reduce fraud. Walmart, for example, uses the IBM Food Trust platform to track food products, significantly reducing the time needed to trace the source of contamination from seven days to just 2.2 seconds (Kamath, 2018). Carrefour has implemented BCT to trace food products like free-range chickens, eggs, and milk, allowing consumers to access detailed product information via QR codes (Kamilaris et al., 2019; CB Insights, 2020).

**Figure 1. Distribution system management in FBSC**

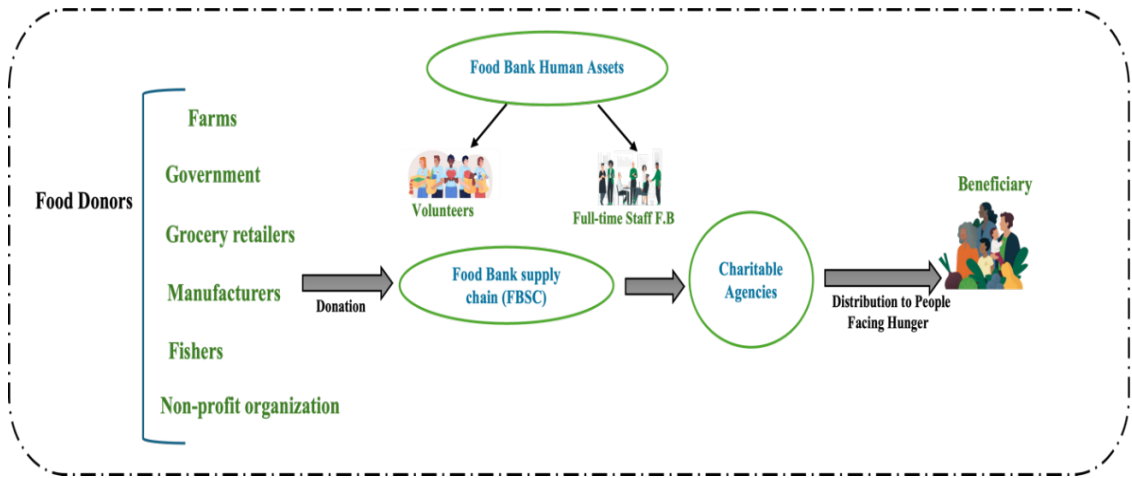

Figure 1. illustrates the flow of resources and relationships within FBSC, focusing on the roles of assorted stakeholders and the flow from donation sources to beneficiaries. The arrows indicate the movement of food and resources from various sources (Government, Grocery Retailers, Manufacturers, Fishers, and Non-profit Organizations) to the FBs. From there, the food and resources are distributed to the beneficiaries. The top part of the diagram shows the two main human resources that FBs rely on: Volunteers and Full-time Staff, who are essential in managing the operations and ensuring the effective distribution of food to beneficiaries. FBs often face transparency, traceability, and efficient management challenges concerning donations and resources. With its immutable and decentralized ledger, blockchain offers a possible fix by ensuring a clear, unchangeable record of every transaction and movement within the SC (Tian, 2016; Kamilaris et al., 2019). Despite the promising benefits of blockchain technology (BCT), food banks encounter numerous barriers to its adoption. To navigate these challenges, this study leverages the Decision-Making Trial and Evaluation Laboratory (DEMATEL) method, a powerful tool for identifying and analyzing the interrelationships among these barriers (Fontela & Gabus, 1976).

This study employs the DEMATEL technique, renowned for its ability to model interdependencies and feedback loops in complex systems (Fontela & Gabus, 1976). By integrating rough set theory with DEMATEL, the approach addresses subjectivity and ambiguity in expert opinions during group decision-making processes (Liou & Tzeng, 2012; Zhang, et al., 2020).

Our contribution to this paper is threefold. First, we explore the adoption of blockchain technology (BCT) and the identification of barriers within the food bank supply chain (FBSC), an area that has not been previously studied in this context. Second, we introduce the Decision-Making Trial and Evaluation Laboratory (DEMATEL) method, a powerful tool for identifying and analyzing the interrelationships among these barriers. Third, we combine the DEMATEL method with rough set theory to prioritize the most significant barriers to BCT adoption in FBSC, offering a structured roadmap for its implementation. By analyzing the causal relationships among barriers, this study enhances the understanding of their interactions and provides actionable insights to improve efficiency, traceability, and transparency in food bank operations. This focus on the practical implementation of research findings is rare in the existing literature on food banks.

Here are the two research questions we are pursuing:

**RQ 1.** What barriers exist for blockchain adoption in the supply chain of food banks?

**RQ 2.** What are the relationships between the barriers?

This paper is structured as follows: Section 2 presents a review of the relevant theoretical background, along with a detailed summary of the contributions of this study. Section 3 outlines the methodology, followed by Section 4, which presents the results. The discussion of these results is provided in Section 5. Finally, Section 6 offers a summary, conclusion, and directions for future research.

## 2. Theoretical background

This section provides an overview of the key concepts underpinning this research, including the adoption of blockchain technology (BCT) in food bank supply chains (FBSCs) and the barriers to its implementation. These insights are summarized in Table 1.

**2.1 Food Bank Supply Chain within Blockchain Technology**

The FAO (2020) highlights its collaboration with various organizations to strengthen research capacities in the least-developed countries. These efforts focus on integrating technology into food systems, with the goal of improving the efficiency and sustainability of food banks. By leveraging innovative technological solutions tailored to local needs, this partnership seeks to address challenges and promote more resilient food systems. In continue, according to the Food and Agriculture Organization (FAO, 2021), food security is achieved when every individual, at all times, has physical, social, and economic access to adequate, safe, and nutritious food that meets their dietary needs and preferences for an active and healthy life. Food security is influenced by a range of interconnected factors, such as conflict, climate change, economic instability, inequality, and poverty. The complex interplay between these factors makes it challenging to fully grasp their collective impact on food insecurity. For further insights, refer to the analysis by Godfray et al. (2010). The integration of blockchain technology into the food industry is expected to generate significant cost savings. By 2024, the combined use of blockchain and Internet of Things (IoT) technologies could save the industry an estimated $31 billion annually. These savings are driven by a reduction in food fraud, enhanced supply chain efficiency, and decreased food waste (Xenos, 2024).

The decentralized digital ledger ensures the immutability of transactions by recording them across a distributed computer network (Nakamoto, 2008). The BCT ledger is distributed across all nodes in the network, allowing each node to access the most up-to-date version (Jellason et al., 2024). Similarly, BCT's reliable traceability system ensures the secure collection and storage of data at every stage of the production process, enhancing safety and enabling effective management of failures or breakdowns (Bonetti et al., 2024; Li et al., 2023; Patelli & Mandrioli, 2020). Thus, digital technologies, such as BCT, can facilitate the sharing of information required to manage and control various issues including timely transfer of foods and ingredients, ensuring the safety of food products throughout the SC, and guaranteeing an adequate shelf life for the end-user. Similarly, digital solutions enable producers to better align supply with demand, helping retailers meet customer needs while preventing oversupply (Annosi et al., 2021; Corallo et al., 2020).

### 3.2 Identifying barriers to the adoption of BCT in FBSCs

*2.3.1 Technological barrier (External views)*

The adoption of new technology is often fraught with challenges, and BCT is no exception. To fully realize the benefits of this technology, it is essential to first overcome these obstacles. All participating parties must thoroughly make sense of these challenges and strategically outline how to address them. This section explored internal barriers within the FBSC. Robust internet connectivity and IT infrastructure are critical for the successful adoption of blockchain. However, in some cases, an organization's IT infrastructure may be inadequate, or access to the necessary technology may be impractical (Kouhizadeh et al., 2021). Studies by Queiroz et al. (2019) and Sabari et al. (2019) identified several barriers to blockchain implementation, including high costs, lack of technological infrastructure, resistance to change, and interoperability issues. These challenges indicate that while blockchain holds great promise, its adoption is far from straightforward. Implementing BCT in the SC could mean lower transaction numbers but higher transaction times. Managing the required environmental and social information is difficult due to the kind, place, and volume of the data. Despite BCT's reputation for having a safe and decentralized structure, recent system breaches and hacks in the cryptocurrency space have raised questions about how vulnerable it is (Yli-Huumo et al., 2016).

*2.3.2 Organizational Policy Barrier*

Another significant barrier is the absence of organizational policy. Organizations must be proactive in defining policies that guide the implementation of new technologies like blockchain. By doing so, they can confidently adopt these technologies and reap their many benefits. Clear guidelines on the proper usage of such technologies and when and where they are most effective give organizations a competitive edge and help them overcome limitations that may hold them back (Saberi et al., 2019). With the right policies in place, organizations can confidently embrace new technologies and take full advantage of their potential (Yadav et al., 2020).

*2.3.3 Infrastructure, facilities, and human resources barrier*

FBs face significant internal challenges due to a lack of essential infrastructure, facilities, and human resources. The limited availability of manpower, inadequate collection and handling operations, and the absence of well-developed distribution infrastructure hinder their day-to-day operations. These resource constraints create obstacles in effectively managing food donations and reaching those in need (Dubey et al., 2022; Hecht et al., 2019).

A report by the Food Research & Action Center (FRAC) highlights that many FBs struggle with inadequate staffing levels and high turnover rates. This can lead to inconsistent service and operational inefficiencies (FRAC, 2022). According to research published in the Journal of Hunger & Environmental Nutrition, training gaps and a lack of specialized skills among staff can affect the management of food resources and overall operational effectiveness; the non-profit sector, including FBs, often faces challenges in maintaining a well-trained workforce. (Hunger & Environmental Nutrition, 2021). According to a study by Feeding America, many FBs operate in facilities that are either too small or poorly equipped to handle the volume of food they need to distribute (Feeding America, 2020). The scarcity of these resources acts as a major barrier, making it tough for them to operate effectively. Adopting new technology systems could be a valuable solution to overcome these barriers. (Millar et al., 2020; Hermsdorf et al.,2017).

*2.3.4 Financial Resources Barrier*

A major challenge within the food banking sector is the persistent lack of financial resources. Food banks rely on surplus food, labor, funding, and various other resources to operate effectively. However, securing adequate funding remains a critical obstacle to their expansion and day-to-day operations. Kouhizade et al. (2021) highlight that insufficient financial support is one of the most significant issues faced by food rescue organizations, directly limiting their ability to function and grow. In Feeding America, financial constraints frequently hinder food banks from expanding programs or introducing new services, such as nutrition education or emergency assistance, thereby limiting their overall impact (Feeding America, 2020). Organisations incur fees for information collecting through FBs and system conversion. Moreover, implementing sustainable practices is expensive. Companies have limited money to implement this technology (Tarasuk et al., 2005; Frasz et al.,2015).

*2.3.5 regulation and legislation about donated food barrier (external views)*

External barriers in the FBSC include regulatory and legislative challenges related to donated food. While donated food is crucial for providing sustenance to those in need, issues such as low nutritional value, limited shelf life, and uncertain supply chains must be addressed. Immediate action is needed to improve the quality and accessibility of donated food. This can be achieved by identifying and developing more effective strategies, such as implementing BCT to enhance traceability and establish a reliable supply chain system. These innovations will maximize the positive impact of donated food, ensuring that everyone has access to nutritious and healthy food (Millar et al., 2020; Alkaabneh et al., 2021).

*2.3.6 Lack of Awareness Among Volunteers Regarding Food Waste/Loss and the Role of Food Banks*

A significant external barrier in the FBSC is the lack of awareness among volunteers about food waste, loss, and the crucial role food banks play. Many volunteers may have limited knowledge in these areas, and the diverse skill sets they bring can further complicate the situation (Akkerman et al., 2023; Annosi et al., 2021). Many volunteers are unaware of the extent of food waste occurring at various points in the supply chain, from production to distribution. This lack of understanding can lead to inefficiencies in sorting, storage, and distribution processes within food banks. Garrone et al. (2014) found that food banks often rely on volunteer labor, but insufficient training and awareness can result in unintended food waste, undermining the goals of food redistribution efforts. Implementing technologies such as BCT could help address this issue by increasing awareness and improving food recovery initiatives. BCT can enhance transparency and traceability across the food supply chain, optimizing resource allocation and minimizing waste (Dubey et al., 2022; Ardra et al., 2022; Frasz et al., 2015).

*2.3.7 Government support for implementing blockchain technology barrier*

The lack of government support for implementing BCT presents a significant barrier. A study by the Food and Agriculture Organization (FAO) highlights that while BCT can greatly improve transparency and efficiency, its adoption is hindered by high costs and the absence of government incentives, particularly in sectors like food banks (FAO, 2020). The World Economic Forum reports that many food banks are unable to afford the upfront costs of BCT, and without government assistance, these barriers are unlikely to be overcome (WEF, 2022). Similarly, the European Commission emphasizes that government-backed regulatory frameworks are crucial for facilitating BCT adoption by ensuring consistency and addressing legal challenges (European Commission, 2022). Governments face multiple challenges in collecting information through food banking and transitioning to new systems. A major obstacle is the cost of adopting sustainable practices, as many countries lack the financial resources to invest in blockchain technology. Additionally, the lack of standardization, proper tools, and procedures for deploying BCT and monitoring sustainability performance further complicates the process. However, overcoming these challenges could significantly enhance government operations and contribute to a more sustainable future (Frasz et al., 2015; Kouhizadeh et al., 2021; Liu et al., 2020; Tarasuk et al., 2005).

**Table 1. The Framework of Internal and External barriers to blockchain adoption in the FBSC**

| Barriers | Sub-barriers | Description. Literature Content | Sources |
|---|---|---|---|
| I - Internal | 1. Lack of access to technology | IT infrastructure and the internet are crucial resources for the deployment of BCT. Sometimes an organization's IT infrastructure is inadequate, or access to technology is not practicable. | (Kouhizadeh et al., 2021). |
| | 2. Lack of organizational policy | Organizations must be proactive in defining policies that guide the implementation of new technologies like blockchain. By doing so, they can confidently adopt these technologies and reap their many benefits. Clear guidelines on the proper usage of such technologies, when and where they are most effective, give organizations a competitive edge and help them overcome limitations that may hold them back. With the right | (Saberi et al., 2019), (Yadav et al., 2020). |

|  |  | policies in place, organizations can confidently embrace new technologies and take full advantage of their potential. |  |
| --- | --- | --- | --- |
|  | 3. Lack of infrastructure, facilities, and human resources | FBs face notable challenges in their day-to-day operations due to the limited availability of crucial resources like manpower, lack of collection and distribution facilities, handling & operations, lack of well-developed infrastructure. The scarcity of these resources serves as a formidable barrier, making it tough for them to operate effectively. Adopting new technology systems could be useful. | (Dubey & Tanksale 2022); Hecht & Neff ,2019) ; (Millar, et al., 2020) ; Hermsdorf et al., 2017). |
|  | 4.Lack of financial resources | For FBs to operate, they need money, labour, extra food, and a number of other resources. Finding enough money to keep FBs open and growing is one of the largest challenges. claimed that one of the biggest challenges faced by food rescue organizations was obtaining funding, and that this posed a barrier to their operations. Organizations pay for information gathering via FBs and system modifications. Moreover, it costs money to put sustainable methods into effect. Businesses lack the funding necessary to use this technology. | (Tarasuk, & Eakin, 2005); (Frasz, et al., 2015). |
| E - External | 5.lack of regulation and legislation pertaining to donated food. | Donated food is a critical resource for providing sustenance to those in need, but it's essential to recognize the challenges that come with it. The low nutritional value, limited shelf life, and uncertain SCs in FBs are unacceptable, and we must take immediate action to address them. We need to identify, develop, and traceability more effective strategies to improve the quality and accessibility of donated food, such as implementing innovative technologies and techniques and establishing a reliable supply chain system. This will enhance the positive impact of donated food to help make certainthat every individual has access to healthy and nutritious food. | (Millar, et al., 2020) ; (Alkaabneh, et al., 2021). |
|  | 6.Lack of awareness among volunteers regarding food waste/loss and food bank roles | Raising awareness among volunteers about the issue of food recovery and food waste in food banks is crucial, as their knowledge on this subject may be limited. The skills of volunteers are diverse and uncertain. It's critical to note that implementing digital processes can significantly decrease food waste and improve food recovery. | (Annosi&Kostoula,2021) ;(Akkerman et al., 2023). Dubey & Tanksale, 2022) ; (Ardra, & Barua, 2022); (Frasz, et al., (2015). |
|  | 7.Lack of government support for implementing blockchain technology | The government faces various obstacles in the process of collecting information through food banking and transitioning to a new system. One of the significant challenges is the cost associated with adopting sustainable practices. Due to limited financial resources, many countries are unable to invest in BCT. Another obstacle is the lack of standardisation and suitable tools, criteria, and techniques for implementing BCT and assessing sustainability output in the government. However, by overcoming these challenges, the government can enhance its operations and assist in  to a more sustainable future. | Kouhizadeh, et al., 2021). Tarasuk, & Eakin, 2005). (Frasz, et al., (2015). |

The next section outlines the methodology used in the study to explore these barriers in depth.

## 3. Methods and materials

We began by identifying the key "barriers" to adopting BCT in the FBSC through a structured literature review. This process revealed two main categories and seven sub-barriers (Table 1). Building on these findings, we designed a survey targeting a selected group of experts from both academia and industry to examine the cause-and-effect relationships among the identified barriers. This survey was approved for data collection and analysis by the Department of Economic Studies at the University of "G. d'Annunzio" Chieti-Pescara, Italy. We invited thirteen experts from one of Italy's food bank networks, Banco Alimentare, including both internal and external stakeholders. These practitioners, primarily in consulting and leadership roles such as managers and consultants, were actively involved in the FBSC. Indeed, Banco Alimentare has established a robust and efficient network across Italy, relying on regional food banks to manage food collection, distribution, and storage for those in need. The organization also plays a vital role in reducing food waste through strategic partnerships with businesses and nonprofits. Additionally, it collaborates with government agencies and European Union initiatives, such as the European Food Banks Federation (FEBA), to extend its reach and impact, feeding millions annually. The network includes 21 regional food banks that operate independently (Banco Alimentare Onlus, 2020, Annual Report; Lambie-Mumford & Dowler, 2014). We also engaged eight academic respondents, all active researchers in the fields of food industry, SCM, and BCT. These participants, including doctoral students, researchers, and professors. Table 2 provides further details on the academic and practitioner respondents. Finally, respondents assessed the influence and impact of each barrier on the others individually, using a Likert scale with five levels: no influence (0), low influence (1), medium influence (2), high influence (3), and very high influence (4). To analyze the data, we applied the rough DEMATEL method and this approach had two primary objectives: First, it aimed to identify the perceived importance of each individual barrier (prominence). Secondly, it sought to understand the perceived interrelationships among these barriers (relations).

**Table 2:** Information on study respondents

| No | Practitioner /Academic | Role | Department/Organization Category |
|----|------------------------|------|----------------------------------|
| 1  | Practitioner | Reporter in Food Bank Supply chain | The Italian Food Bank Netwrk |
| 2  | Practitioner | Food Bank's Executive Director | The Italian Food Bank Network |
| 3  | Practitioner | Advisor (member of the Board of Directors) of the Food Bank) | The Italian Food Bank Network |
| 4  | Practitioner | Technology, IT aspects Expert | The Italian Food Bank Network |
| 5  | Practitioner | Marketing and sales | The Italian Food Bank Network |
| 6  | Practitioner | Advisor for food pantry | The Italian Food Bank Network |
| 7  | Practitioner | Volunteer at the Food Bank | The Italian Food Bank Network |
| 8  | Practitioner | Advisor for Food pantry | The Italian Food Bank Network |
| 9  | Practitioner | Administrative Manager | The Italian Food Bank Network |
| 10 | Practitioner | Administrative employee | The Italian Food Bank Network |
| 11 | Practitioner | Director of food bank | The Italian Food Bank Network |
| 12 | Practitioner | Technology, IT aspects Expert | The Italian Food Bank Network |
| 13 | Practitioner | Advisor for Food pantry | The Italian Food Bank Network |

| 14 | Academic | Professor | Center of Philanthropy & Nonprofit Innovation |
|---|---|---|---|
| 15 | Academic | Associate Professor | Economic |
| 16 | Academic | Doctoral Condidate | Logistics |
| 17 | Academic | Doctoral Student | Operations Management |
| 18 | Academic | Doctoral Student | Economic |
| 19 | Academic | Doctoral student | Supply chain |
| 20 | Academic | Phd Scholar | Business Sciences |
| 21 | Academic | Master Student | Food Industry |

## 3.1. DEMATEL methodology

Analysing interconnected barriers can be challenging, but DEMATEL offers a robust framework for conceptualising cause-and-effect relationships within complex systems (Fontela and Gabus, 1976). It also enables a visual representation of direct and indirect correlations among system components while analysing the causal dependency structure (Song and Cao, 2017). It identifies key variables, both independent and dependent, and quantifies the interactions in a cause-and-effect network (Fu et al., 2012; Lee et al., 2010). Unlike other methods used in supply chain studies to organise and assess a variety of defined obstacles or factors, such as the analytic hierarchy process (AHP) (Saaty, 1988) or interpretive structural modelling (ISM) (Mandal and Deshmukh, 1994), DEMATEL overcomes several limitations. While ISM cannot account for the overall influence of each barrier, AHP ignores the interactions between them. DEMATEL's strength lies in its ability to model these interdependencies and feedback loops, providing a more comprehensive analysis (Biswas and Gupta, 2019; Tzeng et al., 2007). Also, unlike methods such as Failure Mode and Effects Analysis (FMEA), which typically treat barriers as independent factors, DEMATEL explicitly models the interdependencies, making it more suitable for assessing the relationships among multiple barriers (Fontela and Gabus, 1976). This paper combines the rough theory with DEMATEL to collect data to address disagreements, individuality, and ambiguity in judgements by experts during group decision making processes. This method can lead to more reliable and accurate results (Mao et al., 2020; Shojaei et al., 2023). Figure 1 illustrates the research methodology process. The steps are now explained.

**Figure 1.** Research methodology process

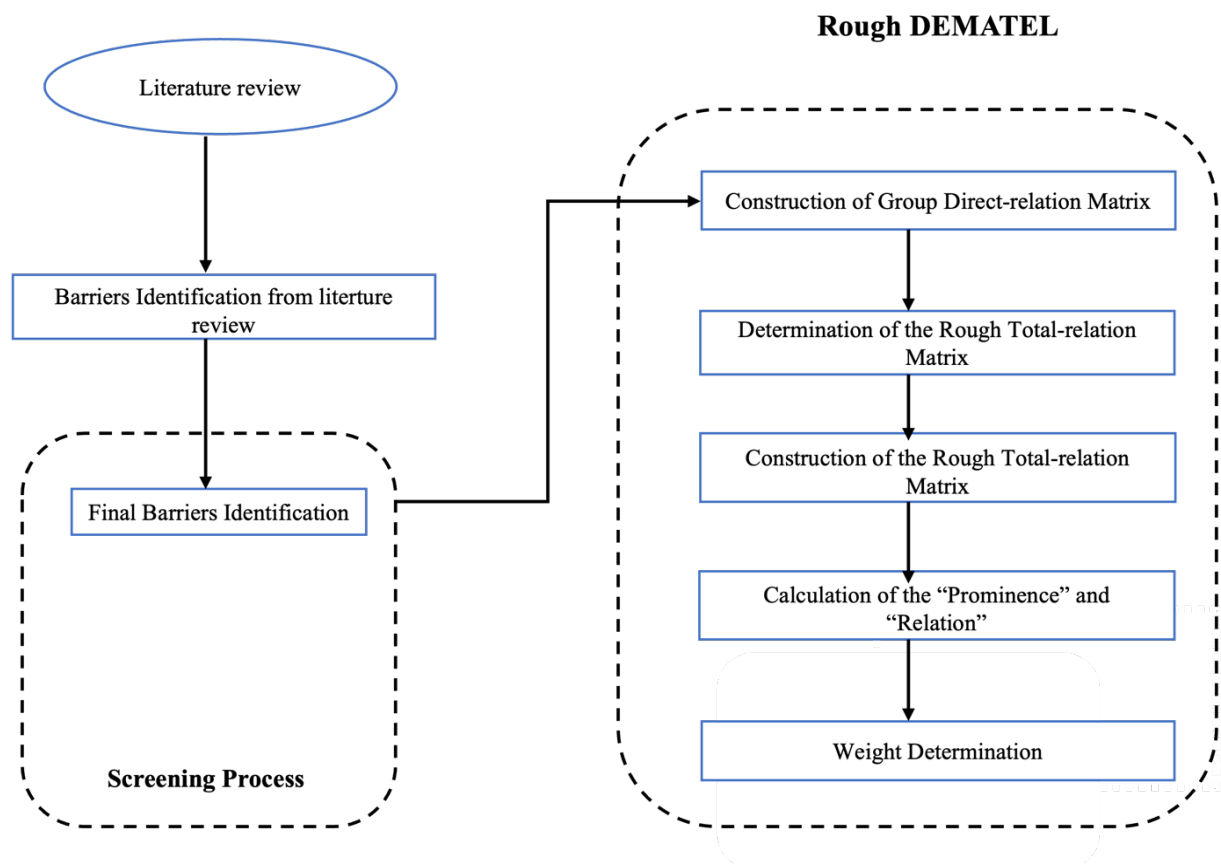

The four steps followed are outlined in the Appendix File.

## 3.2 Rough DEMATEL

Since its introduction by Pawlak in 1982, the rough model theory has developed into a mathematical studies technique that is becoming more and more linked with decision-making approaches. It has been used in conjunction with several key decision-making method, like the Best-Worst method (Haqbin et al., 2021; Rezaei, 2015). The approach for sorting preferences based on likeness to the best alternative The Technique for Order Preference by Similarity to Ideal Solution (TOPSIS) (Chakraborty, 2022; Song et al., 2014), rough Analytical Hierarchy Process - AHP (Gigović et al., 2017), and rough Analytical Network Process - ANP (Li and Wang, 2018). In group decision-making processes, rough theory is particularly useful for aggregating group information to address conflicts, subjectivity, and ambiguity in judgments (Mao et al., 2020), resulting in more accurate and dependable results. The steps of Song and Cao's (2017) rough DEMATEL approach are outlined in the Supplementary File. Mao et al.,

(2020) argues that rough theory helps decrease the ambiguity and subjectivity that result from differing opinions during a group decision-making process by combining group data. The results are explained in the next section.

## 4. Results

### 4.1 Direct-Relation and Normalisation Matrices

Tables A1 and A2, in the supplementary file, show the decision-maker's direct-relation matrix in addition to the group direct-relation matrix and the formulae were subsequently utilised to convert the opinions into approximated numbers. Similarly, the approximate numerical forms for the remaining judgements in the group direct-relation matrix were determined. Table A1 quantifies the direct influence of one criterion on another within the group. This matrix was subsequently normalised (Table A3) to ensure consistency and comparability across criteria, enabling a balanced evaluation of their relationships. [The normalisation process refines the data, allowing for a more accurate analysis of the interactions and dependencies among the elements in the decision-making framework]. In Table A3 the values are scaled to ensure that the total influence of any barrier (the sum of a row or column) does not exceed a value of 1. [The normalisation is achieved by dividing each value by the largest row sum in the initial unnormalised matrix, ensuring consistency in scale and facilitating comparability across barriers].

### 4.2 Total-Relation Matrix and Influence Analysis

Table A4 presents the rough total-relation matrix, showcasing both the upper and lower normalisation values calculated using the MMULT matrix approach. This matrix effectively captures the interrelationships among criteria, providing a comprehensive view of their influence on each other. The upper and lower normalisation helps to standardise the data, ensuring consistency and comparability while highlighting the range of potential values for each criterion within the rough set framework. It can be acquired by using the formula (34)–(36).

After constructing the approximate total-relation matrix T, the sums of the columns $y_j$ and rows $x_i$ were calculated (Table A5). These sums provide valuable insights into the overall 'influence' dynamics among the criteria. Specifically, the row sums represent the degree to which each criterion influences others, while the column sums indicate the extent to which each criterion is influenced by others. By identifying the most influential and dependent criteria, this analysis enhances our understanding of their roles and interconnections in the decision-making process. The "Prominence" $m_i$ and "Relation" $n_i$ are basically determined by taking the total of the columns $Y_j$ are converted into crisp values ($X_i$ $Y_j$). After obtaining the rough total-relation matrix T, the sum of rows $X_i$ and the sum of columns $Y_j$ (see Table S5) are calculated with the formulae (37), (38).

### 4.3 Prominence and Relation of Barriers

Table 3 shows that I1 (3.6135) stands out as the most impactful and prominent barrier, with a strong positive relationship between *X* and *Y* (0.1821). This indicates that addressing this barrier may have a significant and reinforcing effect on the system. I2 and I3 also have high values for both Final Crisp Value and Prominence, and their positive relationships suggest that addressing them may also lead to improvements in the system. These barriers include (lack of access to technology, Lack of organizational policy, and Lack of infrastructure, facilities, and human resources) making it a cause despite its limited overall significance, indicating its active role in influencing other barriers. In contrast, E1 has the strongest negative relationship (-0.3539), This suggests that efforts to address this barrier require a balancing act, and then E2 and E3 also show negative relationships, but their negative impacts are weaker compared to E1.

**Table 3.** The Prominence and Relation values

| Barriers | Final Crisp Value | | Prominence ($m_i$) | Relation ($n_i$) |
|---|---|---|---|---|
| | X | Y | X+Y | X–Y |
| I1 | 3,6135 | 3,4314 | 7,0448 | 0,1821 |
| I2 | 3,4416 | 3,4031 | 6,8447 | 0,0385 |
| I3 | 3,3429 | 3,1950 | 6,5379 | 0,1479 |
| I4 | 3,1392 | 3,1560 | 6,2952 | **-0,0169** |
| E1 | 2,8362 | 3,1900 | 6,0262 | **-0,3539** |
| E2 | 2,9834 | 3,2142 | 6,1976 | **-0,2308** |
| E3 | 3,2453 | 3,3505 | 6,5958 | **-0,1052** |

### 4.4 Normalised Weights and Ranking of Barriers

Table 4 identifies the importance ranking and normalised weights of the barriers. This showed the top five ranked by order of importance as (I1) Lack of organizational access to technology and Lack of organizational policy, (I2) and lack of governance support for BCT implementation (E3) and moreover, the factors cited of least importance were (E1) lack of regulation and legislation about donated food , (E2) lack of awareness among volunteers regarding food waste/loss and bank roles and (I4) lack of financial resources.

I1 and I2 stand out as the most critical barriers, with the highest $\omega_i$ and $W_i$ values. These barriers should be prioritized in any intervention, as they have the greatest overall impact and relative importance. I3 is also significant, ranked 4th in terms of ωi, but slightly less impactful than I1, and I2. E1 is ranked 7th, with the lowest $W_i$ value (0.1325), suggesting it is the least influential barrier. While it may still require attention, it is less critical than the others. I4, E2, and E3 are ranked in the middle (5th, 6th, and 3rd, respectively). These barriers should not be ignored, but they are less critical than I1 and I2. The data highlights that I1, I2, and I3 are the most critical barriers to address, given their high $\omega_i$ and $W_i$ values. E1 is the least significant and should be considered a lower priority. I4, E2, and E3 fall in the middle, requiring attention but not as urgently as the top-ranked barriers. Figure 2 visualises the normalised weights of the barriers (I1, I2, I3, I4, E1, E2, E3), with the height of each orange bar reflecting its relative importance. The exact weights, displayed above each bar, provide a quantitative measure of each barrier's significance within this study, enhancing clarity and interpretation. Notably, barriers I1, I2, E3, emerge as the most critical, indicating their dominant influence and priority in addressing the challenges highlighted in this study. In contrast, barriers such as E1 (0.1325) and E2 (0.1361) demonstrate lower normalised weights, suggesting a reduced impact on the overall system.

**Table 4.** The importance and the normalised weights of the barriers considered

| Barriers | $\omega_i$ | $W_i$ | Ranking |
|---|---|---|---|
| I1 | 7,047184 | 0,1547 | 1 |
| I2 | 6,844819 | 0,1502 | 2 |
| I3 | 6,539578 | 0,1435 | 4 |
| I4 | 6,295243 | 0,1382 | 5 |
| E1 | 6,036575 | 0,1325 | 7 |
| E2 | 6,201863 | 0,1361 | 6 |
| E3 | 6,596673 | 0,1448 | 3 |

**Figure 2.** Weights diagram

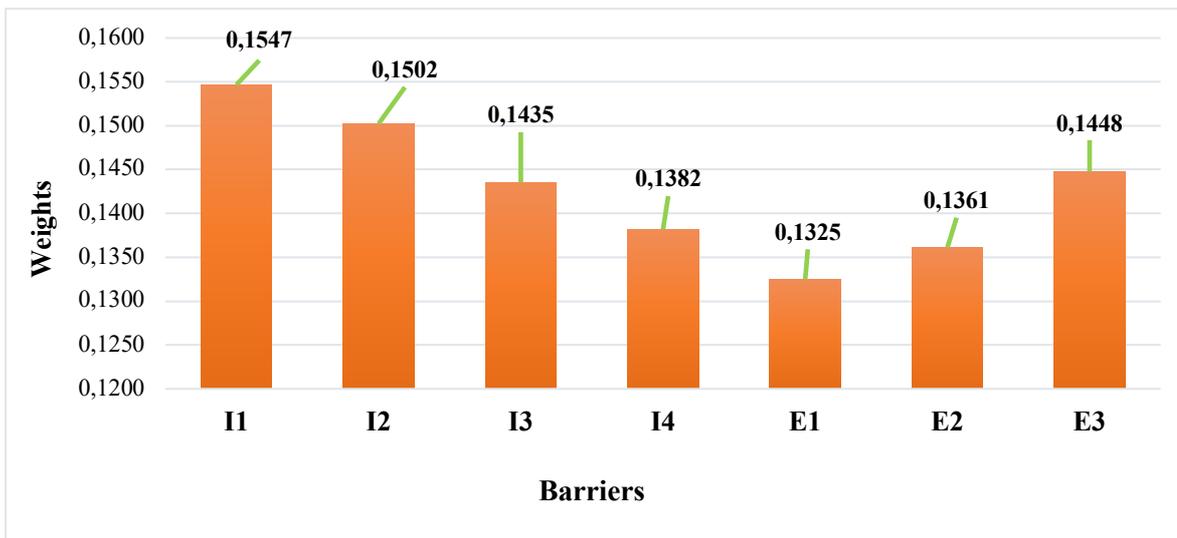

### 4.5 Threshold Analysis and Network Visualisation

Following the approach of Mao *et al.,* (2020), a threshold q was determined in order to identify the significant relationships between the barriers in (T*), The values 2.1475 and 0.2566 represent the mean and standard deviation, respectively, of all components in T*. Furthermore, Threshold T* is used to simplify the direct influence matrix by eliminating less significant or weaker influences, keeping only the stronger relationships. After calculating the average influence matrix (based on multiple expert inputs), a threshold is set to retain only influences that are greater than or equal to this threshold. T* serves as a filter to eliminate weaker connections in the direct influence matrix, focusing the analysis on the more significant relationships. The choice of T* is critical, as it directly affects the complexity and clarity of the resulting model (Tzeng and Huang, 2011).

In Figure 3 the nodes represent the barriers, and the edges indicate significant connections, with their intensity reflected in the thickness and dark orange color of the lines. This visualisation highlights more interactions and provides a broader view of the relationships between barriers. It clarifies the most critical interactions, providing a clear and comprehensive understanding of the relationships between the barriers. the network graph illustrating the significant relationships between the barriers (I1, I2, I3, I4, E1, E2, and E3) with a threshold of 0.5. The nodes represent the barriers, and the edges reflect the relationships exceeding the threshold. The intensity of the relationships is depicted through the thickness and color of the edges, with darker and thicker edges indicating stronger relationships. This visualization helps clarify the interactions and dependencies between the barriers. In contrast, the T2 shows no interaction, reflecting its relatively weak connection with other barriers. In contrast, the E1 and E2 shows no interaction, reflecting its relatively weak connection with other barriers.

**Figure 3.** Significant relationships after the threshold analysis

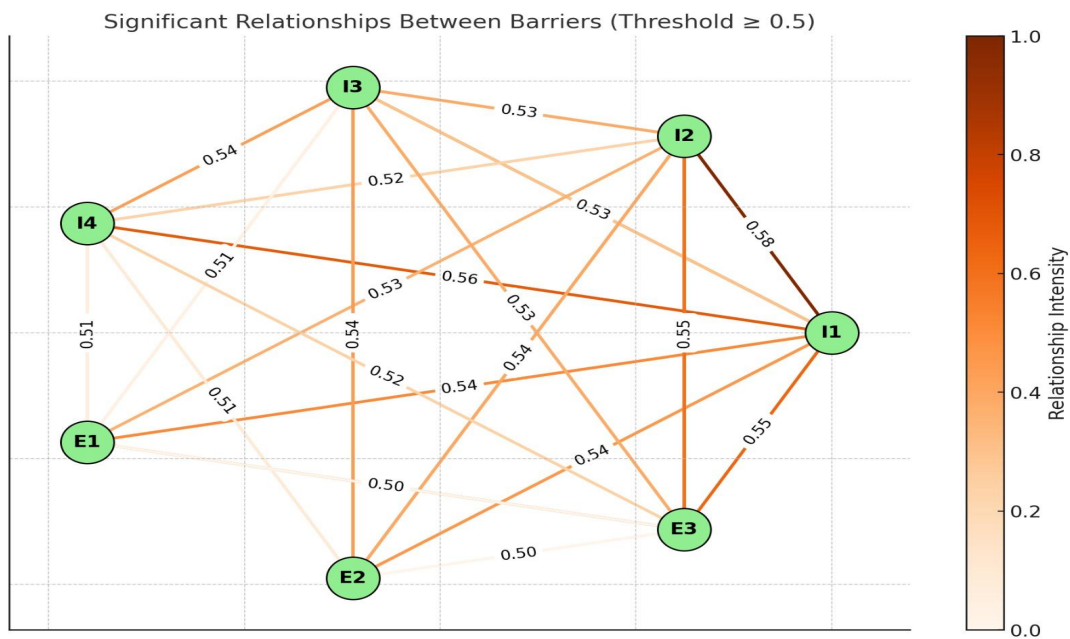

## 4.6 Strategic Insights from the Causal Diagram

The causal diagram (Relation-Prominence) highlights the relationships and significance of the barriers (I1, I2, I3, I4, E1, E2, E3) within the system. Barriers with high prominence, such as I1, I2, and I3, play a critical role in shaping the system's performance. Among these, I1 stands out as the most prominent barrier with a strong positive relation, indicating that addressing it will have a substantial and reinforcing impact on the system. Similarly, I2 and I3 also exhibit high prominence and positive relations, suggesting that resolving these barriers will lead to significant improvements. In contrast, I4 has moderate prominence and a near-neutral relation, indicating a less critical role compared to the top three barriers. On the other hand, barriers E1, E2, and E3 show negative relations, indicating potential challenges when addressing them. E1, in particular, has the strongest negative relation, suggesting that efforts to resolve this barrier may lead to opposing effects, requiring a cautious and balanced approach. While E2 and E3 also exhibit negative relations, their impacts are less severe than E1 but still demand careful consideration. Overall, the diagram suggests prioritizing the resolution of I1, I2, and I3 to achieve the most significant positive impact while adopting strategic and cautious approaches to manage barriers with negative relations, such as E1, E2, and E3. (See Figure 4)

**Figure 4.** The Causal Diagram (Relation-Prominence)

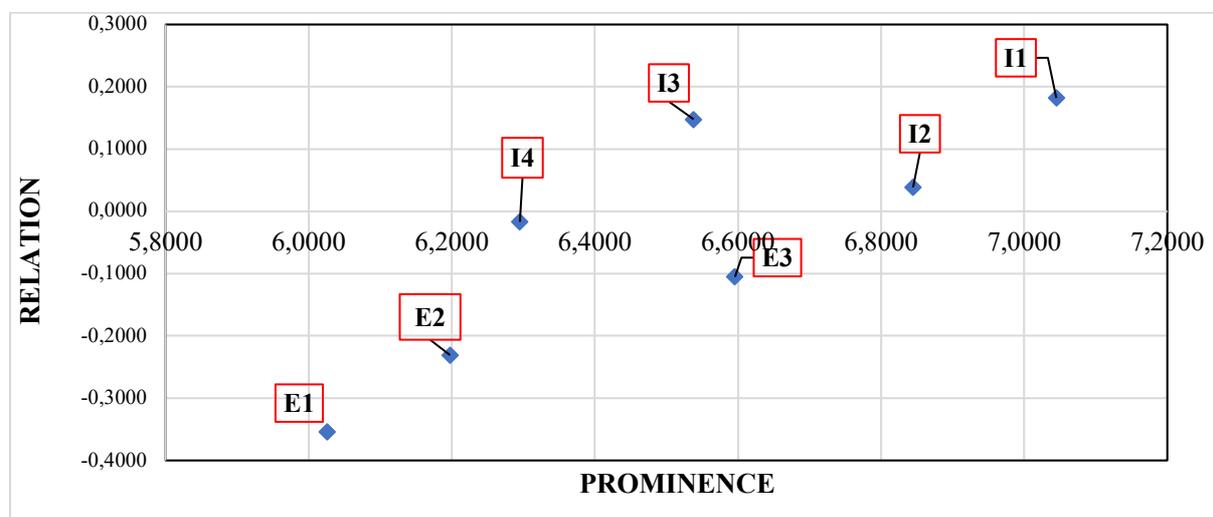

## 5. Discussion

The relation value groups obstacles into cause or effect groups, as described in the steps of the rough DEMATEL process. The barrier in question belonged to the cause group in circumstances with a positive relation value. Otherwise, the relation value was negative, which is the effect group's barrier that applies. Therefore, the functionality of the obstacles depends on the operation of the other barriers. The findings revealed that Lack of access to technology (I1), Lack of organizational policy (I2), and Lack of infrastructure, facilities, and human resources (I3) were the primary drivers of other barriers, as they had positive relation values (ni). Of these, Lack of infrastructure, facilities, and human resources management (I3) emerged as the most critical factor, with a prominence value of 0.1479, indicating its significant influence. Table 4 illustrates the normalized weights and the importance barriers. "lack of financial resources" (I4), "lack of regulation and legislation about donated food" (E1), "lack of awareness among volunteers regarding food waste/loss and bank roles" (E2), and "lack of government support for implementing blockchain technology" (E3) were the causal barrier of least importance with the lowest prominence value (mi).

As illustrated in Figure 5, the most significant issue identified is the lack of access to technology (I1). Technological barriers can influence organizational policy challenges, which in turn affect SC barriers. Organizational barriers often serve as a bridge between technological and SC challenges. For example, the immaturity of BCT as a critical technological concern may create hesitation among stakeholders or managers, potentially impacting their commitment to and support for its implementation within FBSc (Iansiti & Lakhani, 2017; Saberi et al., 2019). Technological and FBSc related barriers emerge as the most critical categories requiring focused attention. Addressing technological barriers (I1), as an internal challenge, is a priority for mitigating SC obstacles and advancing BCT adoption in the FBSc. Tackling these technological challenges within organizational policies can significantly reduce broader FBSc barriers (Tian, 2017). Furthermore, limited awareness among volunteers about food waste and the vital role of FBs remains a crucial external barrier that also requires immediate action.

Nevertheless, the lack of expertise and knowledge makes effect hesitate to change to new systems. This leads to a relationship mediated between a lack of management commitment and support. The adoption of BCT can help them raise updated awareness among volunteers about all issues in FBs. (Alexander & Smaje, 2008; Davis, 1989). Additionally, this barrier has a low relative prominence compared to the other barriers, however, this could be owing to a lack of influence over this barrier (E2). Nonetheless, certain studies in the FBSC literature may be somewhat compared to the findings of the current investigation. For example, food banks faced multiple unprecedented hurdles as a result of the pandemic and were not immune to the effects of the COVID-19 issue. Due to the various social restrictions imposed, many FBs and their forefront organizations faced significant challenges, a shortage of human resources like volunteers, both essential to the daily operations of FBs and their frontline partners, created a substantial strain on their ability to function effectively.

Moreover, several frontline organizations that were collecting food from the Polish FB were forced to temporarily close due to tight regulations and a shortage of volunteers. As a result, the Polish FB redistributed less food in the spring of 2020, because of a limit of food but because of the organizations' lack of personnel to redistribute it (Capodistrias, 2022). On the other hand, governments encounter several challenges when collecting information through FBs and transitioning to a new system. One major challenge is the high cost of adopting sustainable practices, which many countries struggle to afford due to limited financial resources, making investment in BCT difficult.

Further, the lack of standardization, as well as the unavailability of relevant methods, tools, measurements, and procedures for deploying BCT and monitoring sustainability functioning, is a substantial challenge. However, by overcoming these obstacles, the government may improve its operations and make a significant contribution to a more sustainable future (Kouhizadeh et al., 2021). Nevertheless, BCT enables more effective monitoring and enforcement of food safety standards by supplying a clear, Tamper-proof record of food transactions. Policymakers can use these records to ensure that FBs comply with regulations, potentially reducing the incidence of foodborne illnesses and policymakers can also leverage the transparency provided by BCT to encourage the adoption of best practices in food distribution. For example, BCT data can help identify inefficiencies or bottlenecks in the SC, informing policy decisions to improve food security (Casino et al., 2019). FBs can use BCT to generate accurate and detailed reports for donors, showcasing the impact of their contributions. This capability is particularly important for attracting and retaining large donors, including corporate sponsors and institutional funders (Saberi et al., 2019).

Likewise, Policymakers can introduce tax incentives for donations made through BCT-verified platforms, ensuring that contributions are tracked and used efficiently. This could encourage more substantial and more frequent donations to FBs (Zyskind et al., 2015). The use of BCT in FBs can improve sustainability by enabling precise tracking of food donations and distributions, ultimately reducing waste and improving resource allocation. Saberi et al. (2019) discuss how BCT can enhance resource management and support sustainability in non-profit settings. If the results of implementing BCT in FBs show reduced waste and better management of food resources, it would not only reinforce existing sustainability theories but also extend them by demonstrating how BCT can serve as a powerful tool for achieving sustainability goals in non-profit settings.

## 5.1. Managerial implications

BCT is perceived as too complex or underdeveloped, leaders may be hesitant to invest resources, thereby slowing down the adoption process. Organizations need to first focus on upskilling their workforce and establishing a clear understanding of BCT's capabilities before attempting widespread adoption (Saberi et al., 2019). Leaders need to align their strategic goals with adoption, ensuring that any technological investments align with long-term organizational objectives (Tian,2017). To successfully implement BCT in FBSC, managers must foster strong partnerships with stakeholders and establish clear agreements on data governance and information-sharing protocols. These measures are crucial for ensuring seamless blockchain integration across the supply chain (Queiroz & Wamba, 2019). Organizational resistance to change, including insufficient managerial support or a reluctance to embrace new technologies, can significantly hinder the successful adoption of BCT in supply chains. Addressing these barriers requires securing leadership commitment, offering comprehensive training, and fostering a culture of innovation. Such efforts not only help overcome internal resistance but also enhance collaboration with supply chain partners an essential factor for the transparent and decentralized functionality of BCT systems (Queiroz & Wamba, 2019; Kshetri, 2018).

To address the lack of Organizational Policy (I2), managers should develop clear policies for data governance, information sharing, and blockchain implementation to align technological adoption with organizational goals. Strong leadership support is essential to drive innovation, ensure strategic alignment, and communicate policies effectively across all levels. For the Lack of Access to Technology (I1), managers must prioritize upskilling employees and volunteers through targeted training programs to enhance technological proficiency. Partnerships with tech providers or non-profits can help access affordable solutions, such as cloud-based platforms or blockchain-as-a-service (BaaS), reducing implementation costs. To overcome the Lack of Infrastructure, Facilities, and Human Resources (I3), managers should streamline operations and invest in critical infrastructure, such as upgraded storage facilities and smart logistics systems, to improve efficiency. Collaborating with government agencies, non-profits, and private stakeholders can secure funding and resources, including manpower, transportation, and distribution support.

## 5.2. Practical implications

The practical implications highlight the need for organizations to prioritize upskilling their workforce and fostering a clear understanding of BCT's capabilities before pursuing widespread adoption (Saberi et al., 2019). Training programs aimed at enhancing technological proficiency among employees and stakeholders are essential for ensuring the effective implementation of blockchain solutions. Additionally, organizations must work collaboratively to establish shared standards and protocols for BCT, facilitating seamless data exchange across platforms and improving overall supply chain efficiency. Nonprofit organizations, manufacturers, government agencies, and charitable institutions should collaborate to promote the development and adoption of BCT. By prioritizing cybersecurity and data protection, the industry can implement best practices that build trust in blockchain solutions. This approach not only reduces risks related to fraud and data breaches but also encourages confidence among stakeholders, driving widespread adoption and ensuring the successful integration of BCT into food distribution systems.

## 6. Conclusion and Future Research Direction

This study aimed to identify and analyze the interrelationships among barriers to adopting BCT in the FBSc. These barriers were categorized into internal and external groups, and the rough group DEMATEL technique was applied to evaluate their interactions. The methodology effectively revealed the relationships between barriers, providing a clear understanding of their influence. By addressing the blockchain technology gap, this study highlights opportunities to enhance the adoption and effectiveness of SC innovations. Furthermore, the establishment of robust organizational policies can create a supportive framework for integrating new technologies. The findings offer valuable insights for practitioners and policymakers to prioritize actions, overcome barriers, and strengthen the efficiency and resilience of supply chains. It should be noted that this study has limitations such as the potential challenge in generalizing the findings across all FBSC. The specific barriers to blockchain adoption identified through the Rough DEMATEL-Based Approach may vary depending on the size, structure, and operational practices of different food banks. As a result, the insights gained might not be universally applicable to all organizations within this sector. Furthermore, Barriers to adoption may differ significantly across various geographic regions, influenced by local regulations, technological infrastructure, and cultural attitudes toward digital innovation. This limitation suggests that the findings may not be entirely relevant or applicable in all contexts. Likewise, it offers novel insights into the integration of BCT within the FBSc, particularly by identifying and analyzing the theoretical adoption barriers. Future research could explore tailored solutions to overcome adoption barriers across different regions and countries, considering variations in technological infrastructure, regulatory frameworks, and cultural attitudes toward BCT. Additionally, studies could investigate how BCT enhances transparency and trust within the FBSC, particularly in tracking donations, financial flows, ensuring food safety, and minimizing waste. Examining the regulatory implications of BCT adoption, with a focus on data privacy, security, and compliance with food safety standards, is another critical avenue for future research.

Besides that, we recommend exploring implementing BCT decision models that align sector-specific needs with suitable design features. Additionally, longitudinal studies could assess the long-term impact of overcoming adoption barriers on key supply chain performance metrics, such as traceability, efficiency, and sustainability. Incorporating advanced methodologies, such as hybrid multi-criteria decision-making approaches or machine learning techniques, could further enhance the precision and depth of barrier analysis in future studies.

# Appendix A:

**Step 1.** Establishing matrix Z, the average initial direct-relation matrix

Consider the system requirements are: B = {B1, B2, B3,…,Bn}. all characteristics are rated on a five-point scale: No Influence (0), Very Low Influence (1), Low Influence (2), High Influence (3), and "Very High Influence" (4). Due to the DEMATEL method's properties, the diagonal members of the matrix have a value of 0. The components of matrix Z is computed outlined below:

$$z_{ij} = (1/H) \sum_{k=1}^{H} x_{ij}^{k} \qquad (1)$$

Where $x_{ij}$ is the non-negative valued matrix for the *k*-th answer, $x_{ij}^{k}$ is the degree of influence of criterion i on criterion j with regard to the *k*-th answer, H is the total amount of experts, n is the number of criteria, and $x_{ij}^{k}$ is the degree of influence supplied by criterion i to criterion j. where $x_{ij}^{k}$ is the degree to which the criterion i influences the criterion j with regard to the *k*-th response, H is the number of experts, n is the number of criteria, Xk is the not negative matrix for the *k*-th response, and $x_{ij}$ is the degree to which the criterion i influences the criterion j. The first direct relation matrix (Zn×n) is then built in the manner described below:

$$Z = \begin{bmatrix} z_{11} & z_{12} & \cdots & z_{1n} \\ z_{21} & z_{22} & \cdots & z_{2n} \\ \vdots & \vdots & \ddots & \vdots \\ z_{n1} & z_{n2} & \cdots & z_{nn} \end{bmatrix} \qquad (2)$$

Where $z_{ij}$ denotes the strength of indicator i's effect over a measure j.

**Step 2.** Making the normalised original direct-relation matrix (Matrix D). The equation that follows is used to calculate a normalised initial direct-relation matrix, or matrix-D:

$$S = 1/(\max_{1<i<n} \sum_{j=1}^{n} z_{ij}) \qquad (3)$$

$$D = Z \times S \qquad (4)$$

**Step 3.** Making a factor's total-influence matrix (Matrix T)

The following formula is utilised to calculate a factor's total-influence matrix, or matrix T:

$$T = D(I - D)^{-1} \qquad (5)$$

So that, **(I)** refer to an identity matrix.

**Step 4.** Computing the values of *D, R, R+D,* and *R-D*

Below describes how the numbers are computed: D is the sum of the matrix *T*'s columns, $D_j$ indicates the other criteria's both direct and indirect effect on criterion j, R is the sum of the matrix *T*'s rows, and $R_i$ indicates the both the direct and indirect impact of criterion (i) on the other criteria. subsequently, $R_i$-$D_j$ illustrates the net impact of criterion I, while $R_i$+$D_j$ shows the significance of criterion (i).

**Proposition 1.** If A represent a set of n classes {A1, A2,..., An} that cover each object in U, P be a random object from U, and U be the entire universe including all objects. The classes to that the item belongs are represented with ∀P ∈ U, Ak ∈ R, and $1 \leq k \leq n$ if the categories are classified as {A1 < A2 < … < An}. The category Ak's lower and upper approximations, in addition to its boundary region, are defined as:

$$\underline{Apr}\,(A_k) = \{P \in U | R(P) \leq A_k\} \qquad (6)$$

$$\overline{Apr}\,(A_k) = \{P \in U | R(P) \geq A_k\} \qquad (7)$$

$$(A_k) = \{P \in U | R(P) \neq A_k\} = \{P \in U | R(P) > A\} \cup \{P \in U | R(P) < A\} \qquad (8)$$

**Proposition 2.** Given the corresponding upper and lower the boundaries, Ak might be defined as an approximate value RN ($A_k$):

$$\underline{Lim}\,(A_k) = \frac{1}{M_L} \sum \{P \in \underline{Apr}\,(A_k)\} \qquad (9)$$

$$\overline{Lim}\,(A_k) = \frac{1}{M_U} \sum \{P \in \overline{Apr}\,(A_k)\} \qquad (10)$$

$$RN\,(A_k) = [\underline{Lim}\,(A_k), \overline{Lim}\,(A_k)] \qquad (11)$$

Here, $M_L$ and $M_U$ have been the total number of objects in *Apr*, correspondingly.

**Proposition 3.** The difference across both the lower and upper limits is expressed through the rough border intervals. $\overline{Lim}\,(A_k) - \underline{Lim}\,(A_k) = RN\,(A_k)$

**Proposition 4.** The following mathematical operations are carried out for the two rough numbers, RN (α) = [$\underline{Lim}$ (α), $\overline{Lim}$ (α)] and RN (β) = [$\underline{Lim}$ (β), $\overline{Lim}$ (β)]

RN (α) + RN (β) = [$\underline{Lim}$ (α) + $\underline{Lim}$ (β), $\overline{Lim}$ (α) + $\overline{Lim}$ (β)]  (12)

RN (α) - RN (β) = [$\underline{Lim}$ (α) - $\underline{Lim}$ (β), $\overline{Lim}$ (α) - $\overline{Lim}$ (β)]  (13)

RN (α) × RN (β) = [$\underline{Lim}$ (α) × $\underline{Lim}$ (β), $\overline{Lim}$ (α) × $\overline{Lim}$ (β)]  (14)

$\frac{RN\ (\alpha)}{RN\ (\beta)} = [\frac{\underline{Lim}\ (\alpha)}{\underline{Lim}\ (\beta)}, \frac{\overline{Lim}\ (\alpha)}{\overline{Lim}\ (\beta)}]$  (15)

μ × RN (α) = [μ × $\underline{Lim}$ (α), μ × $\overline{Lim}$ (α)]  (16)

**Proposition 5.** Utilise one of the following equations to convert a rough number into a crisp value:

$$W_k^e = \frac{\tilde{C}_k^e}{\sum_{k=1}^n \tilde{C}_k^e} \quad (17)$$

$$\tilde{C}_k^e = \min_k\{w_k^{e*L}\} + \chi_k^e \times (\max_k\{w_k^{e*U}\} - \min_k\{w_k^{e*L}\}), k = 1, 2, \ldots, n \quad (18)$$

$$\chi_k^e = \frac{w_k^{e*L} \times (1 - w_k^{e*L}) + w_k^{e*U} \times w_k^{e*U}}{1 - w_k^{e*L} + w_k^{e*U}} \quad (19)$$

Zhu *et al.,* (2017) define the following terms in relation to rough numbers:

**Step 1:** Formulation of direct-relation matrix for crisp group

Based on expert responses the direct-relation matrix $M_1$ of the first decision maker can be produced using Equation 20.

The direct-relation matrix $M_k$ for the *k*-th experts is generated by: The *k*-th expert's direct-relation matrix is $M_k$ produced through the following equation:

$$M_k = \begin{bmatrix} 0 & r_{12}^k & \cdots & r_{1n}^k \\ r_{21}^k & 0 & \cdots & r_{2n}^k \\ \vdots & \vdots & \ddots & \vdots \\ r_{n1}^k & r_{n2}^k & \cdots & 0 \end{bmatrix}, \quad k = 1, 2, \ldots, m \quad (20)$$

The *k*-th expert's perceptive evaluation is represented by $r^k_{ij}$ for the ith items. wherein $r^k_{ij}$ is the *k*-th expert's exact assessment of the ith element's influence on the *j*-th element, and m, n, and n are the numbers of experts, elements, and experts, respectively. Impact regarding the *j*-th component, wherein m, and n indicate the number of components and experts, respectively. To create the group's direct-relation matrix $\tilde{R}$, utilised the following:

The direct-relation matrix of the group $\tilde{R}$ :

$$\tilde{R} = \begin{bmatrix} 0 & r_{12}^{\sim k} & \cdots & r_{1n}^{\sim k} \\ r_{21}^{\sim k} & 0 & \cdots & r_{2n}^{\sim k} \\ \vdots & \vdots & \ddots & \vdots \\ r_{n1}^{\sim k} & r_{n2}^{\sim k} & \cdots & 0 \end{bmatrix} \text{ where: } r_{ij} = \{r_{ij}^1, r_{ij}^2, \ldots, r_{ij}^k, \ldots, r_{ij}^m\} \quad (21)$$

**Step 2:** Establishment of the rough group's direct relationship matrix

$$\underline{Apr}(r_{ij}^k) = \cup \{P \in U \mid J(P) \leq r_{ij}^k\} \quad (22)$$

$$\overline{Apr}(r_{ij}^k) = \cup \{P \in U \mid J(P) \geq r_{ij}^k\} \quad (23)$$

Assume that there are m types of opinions from experts for every component relation. $J = \{r_{ij}^1, r_{ij}^2, \ldots, r_{ij}^k, \ldots, r_{ij}^m\}$ has been arranged in the subsequent order: $r_{ij}^1 < r_{ij}^2 < \ldots < r_{ij}^k < \ldots < r_{ij}^m$. $P$ is a randomised element of $U$. The following describes the lower and upper approximations of $r_{ij}^k$: the lower approximation illustrates Equation (22) and the upper approximation depicts Equation (23).

The judgement $r_{ij}^k$ may be expressed as a rough number with lower limit $\underline{Lim}(r_{ij}^k)$ and upper limit $\overline{Lim}(r_{ij}^k)$ outlined below:

$$\underline{Lim}(r_{ij}^k) = \frac{\sum_{m=1}^{N_{ijL}} x_{ij}}{N_{ijL}} \quad (24)$$

$$\overline{Lim}(r_{ij}^k) = \frac{\sum_{m=1}^{N_{ijU}} y_{ij}}{N_{ijU}} \quad (25)$$

$x_{ij}$ and $y_{ij}$ indicates both lower and upper approximation for $r_{ij}^k$. The numerical values of objects that are contained in the lower and higher approximations of $r_{ij}^k$, Correspondingly, are denoted by the words $N_{ijL}$ and $N_{ijU}$. In the direct relation matrix $M_k$, all of the crisp judgements $r_{ij}^k$ are converted to rough values (Zhai et al., 2009). $RN(r_{ij}^k)$ is then given can be obtained using Equations. (2)-(5),

$$RN(r_{ij}^k) = [\underline{Lim}(r_{ij}^k), \overline{Lim}(r_{ij}^k)] = [r_{ij}^{kL}, r_{ij}^{kU}] \tag{26}$$

Where the rough number $RN(r_{ij}^k)$ in the direct relation matrix of value k-th has lower and upper bounds, denoted, respectively, by $r_{ij}^{kL}$ and $r_{ij}^{kU}$. The degree of ambiguity is donated by the boundary region's interval. Moreover, the rough number will be more accurate the narrower the boundary region's interval (Song and Cao, 2017). Thus, it is possible to derive an approximate sequence $RN(\tilde{r}_{ij})$ as follows:

$$RN(\tilde{r}_{ij}) = \{[r_{ij}^{1L}, r_{ij}^{1U}], [r_{ij}^{2L}, r_{ij}^{2U}], \ldots, [r_{ij}^{mL}, r_{ij}^{mU}]\} \tag{27}$$

Using rough computation methods, one may compute the average rough interval $\overline{RN(\tilde{r}_{ij})}$:

$$\overline{RN(\tilde{r}_{ij})} = [r_{ij}^L, r_{ij}^U] \tag{28}$$

$$r_{ij}^L = \left(\sum_{k=1}^{m} r_{ij}^{kL}\right)/m \tag{29}$$

$$r_{ij}^U = \left(\sum_{k=1}^{m} r_{ij}^{kU}\right)/m \tag{30}$$

$r_{ij}^L$ and $r_{ij}^U$ are lowers and upper restrictions of rough number $[r_{ij}^L, r_{ij}^U]$ represented, wherein $m$ is the number of judgements.

The next equation represents the process of the rough group direct-relation matrix R able to be constructed as follows:

$$R = [\overline{RN\ (\tilde{r}_{ij}^k)}]_{n \times n} = \begin{bmatrix} [0,0] & [r_{12}^L, r_{12}^U] & \cdots & [r_{1n}^L, r_{1n}^U] \\ [r_{21}^L, r_{12}^U] & [0,0] & \cdots & [r_{2n}^L, r_{2n}^U] \\ \vdots & \vdots & \ddots & \vdots \\ [r_{n1}^L, r_{12}^U] & [r_{n2}^L, r_{12}^U] & \cdots & [0,0] \end{bmatrix} \tag{31}$$

**Step 3:** Make a roughly total-relation matrix

To normalise the element scales into comparable scales, apply the linear scale transformation formula. The normalised rough group direct-relation matrix R' is created by following these steps:

$$R' = [\overline{RN\ (r_{ij})}']_{n \times n} = \begin{bmatrix} \overline{RN\ (\tilde{r}_{11})}' & \overline{RN\ (\tilde{r}_{12})}' & \cdots & \overline{RN\ (\tilde{r}_{1n})}' \\ \overline{RN\ (\tilde{r}_{21})}' & \overline{RN\ (\tilde{r}_{22})}' & \cdots & \overline{RN\ (\tilde{r}_{2n})}' \\ \vdots & \vdots & \ddots & \vdots \\ \overline{RN\ (\tilde{r}_{n1})}' & \overline{RN\ (\tilde{r}_{n2})}' & \cdots & \overline{RN\ (\tilde{r}_{nn})}' \end{bmatrix} \quad (32)$$

Moreover, the rough total-relation matrix (T) can be generated using:

$$\overline{RN\ (r_{ij})}' = \frac{\overline{RN\ (\tilde{r}_{ij})}}{\tau} = [\frac{r_{ij}^L}{\tau}, \frac{r_{ij}^U}{\tau}],\ \tau = \max_{1 \le i \le n}(\sum_{j=1}^{n} r_{ij}^U) \quad (33)$$

The approximate total-relation matrix, $T$, is capable of being created as follows:

$$T = [tij]\ n \times n, \quad (34)$$
$$t_{ij} = [t_{ij}^L, t_{ij}^U], \quad (35)$$
$$T^S = [t_{ij}^s]\ n \times n = R'^S\ (I - R'^S)^{-1},\ s = L, U. \quad (36)$$

In the total-relation matrix $T$, the lower and upper boundaries of the rough range $t_{ij}$ are denoted by $t_{ij}^L$ and $t_{ij}^U$, correspondingly, whereas I is the element of the matrix.

The rough interval's lower and upper limitations, as represented by the total-relation matrix $T$. $t_{ij}$ are denoted by $t_{ij}^L$ and $t_{ij}^U$, respectively, with I as the value of the matrix.

**Step 4: Calculation of the "prominence" and "relation" values**

After getting the preliminary total-relation matrix $T$, we must calculate the sums of rows $X_i$ and columns $Y_j$. The sums of the columns are represented by $x_i$ and $y_j$ in the rough total-relation matrix using the equations below. An approximate total-relation matrix is represented by the sum of rows and columns.

$$X_i = [x_i^L, x_i^U] = [\sum_{j=1}^{n} t_{ij}^L, \sum_{j=1}^{n} t_{ij}^U] \quad (37)$$

$$Y_j = [y_j^L, y_j^U] = [\sum_{i=1}^{n} t_{ij}^L, \sum_{i=1}^{n} t_{ij}^U] \quad (38)$$

$x_i^L$ and $x_i^U$ are the upper and lower limitations of the interval's rough $x_i$. Similarly. $y_j^U$ and $y_j^L$ are the upper and lower restrictions and of the rough interval $Y_i$. To effectively determine the Prominence and Relation, it is necessary to convert the $X_i$ and $Y_j$ into crisp values. A "the roughness" of X is calculated as follows:

(1) Normalisation of data

$$\tilde{x}_i^L = \left(x_i^L - \min_i x_i^L\right) / \Delta_{min}^{max} \quad (39)$$

$$\tilde{x}_i^U = \left(x_i^U - \min_i x_i^L\right)/\Delta_{min}^{max} \tag{40}$$

$$\Delta_{min}^{max} = \max_i x_i^U - \min_i x_i^L \tag{41}$$

Where $\tilde{x}_i^L$ and $\tilde{x}_i^U$ are the normalised from of the $x_i^L$ and $x_i^U$ corresponding.

(2) Identify the overall normalised in crisp value

$$\alpha_i = \frac{\tilde{x}_i^L \times (1 - \tilde{x}_i^L) + \tilde{x}_i^U \times \tilde{x}_i^U}{1 - \tilde{x}_i^L + \tilde{x}_i^U} \tag{42}$$

Computing the final crisp values $x_i$ for $X_i$

$$x_i = \min_i x_i^L + \alpha_i \Delta_{min}^{max} \tag{43}$$

Likewise, we may calculate the final crisp values $y_i$ for $Y_i$..The vector $m_i$, termed Prominence, is created by adding $x_i$ to $y_i$. Similarly, the vector $n_i$ named Relation, is created by deducting $x_i$ to $y_i$.

$$m_i = x_i + y_j, \quad i = j \tag{44}$$
$$n_i = x_i - y_j, \quad i = j \tag{45}$$

The vector $m_i$ indicates prominence, while the vector $n_i$ gauges the relationship. Prominence indicates the importance of the criterion. As an outcome, the higher its prominence value, the more significant the criterion is in relation to the other criteria. Separate the criteria into cause-and-effect groups. When a criterion has a positive relation value, it is assigned to the cause group. In contrast, if the relation value is negative, the criterion is classified as an effect. In such cases, the criteria would be reliant on the activities of the additional requirements. Song and Sakao (2018) highlight that mapping the data set of prominence and relation values yields the causal diagram.

**Step 5: Assess the weight of adoption barriers in the FSC.**

The relevance of the criteria is computed as follows:

$$\omega_i = \sqrt{m_i^2 + n_j^2} \tag{46}$$

Moreover, the significance of the criterion can be normalised as follows.

$$W_i = \frac{\omega_i}{\sum_{1 \leq i \leq n} \omega_i} \tag{47}$$

Based on Song and Cao (2017) each criterion's weight indicates how important it is in relation to the others. Tables S1 and S2 depict the decision-maker's direct-relation matrix in addition to the group direct-relation matrix and formulae were subsequently utilised to convert the opinions into approximated numbers.

$$\underline{Lim}\,(A_k) = \frac{1}{M_L} \sum \{P \in \underline{Apr}\,(A_k)\} \qquad (48)$$

**Table A1.** The first decision maker's direct-relation matrix

| Barriers | I1 | I2 | I3 | I4 | E1 | E2 | E3 |
|---|---|---|---|---|---|---|---|
| **I1** | 0 | 4 | 4 | 0 | 0 | 4 | 0 |
| **I2** | 1 | 0 | 2 | 0 | 0 | 4 | 0 |
| **I3** | 4 | 3 | 0 | 0 | 0 | 4 | 0 |
| **I4** | 4 | 2 | 4 | 0 | 0 | 3 | 0 |
| **E1** | 2 | 2 | 1 | 0 | 0 | 3 | 3 |
| **E2** | 4 | 4 | 4 | 0 | 0 | 0 | 0 |
| **E3** | 3 | 3 | 2 | 0 | 0 | 3 | 0 |

**Table A2.** The rough group direct-relation matrix

| Barriers | I1 | I2 | I3 | I4 | E1 | E2 | E3 |
|---|---|---|---|---|---|---|---|
| I1 | [0.0000,0.0000] | [1.8186,3.2600] | [17305,3.2076] | [1.3424,3.0724] | [1.6838,3.3281] | [1.4919,3.2486] | [1.2090,2.8690] |
| I2 | [1.8029,3.5333] | [0.0000,0.0000] | [1.4219,2.9943] | [1.3329,2.8724] | [1.2576,2.8814] | [1.5152,3.1843] | [1.5443,3.3905] |
| I3 | [1.7881,3.1119] | [1.6967,3.2381] | [0.0000,0.0000] | [1.6510,3.1676] | [0.9614,2.4419] | [1.8029,3.0776] | [1.3305,3.0586] |
| I4 | [1.8029,3.5333] | [1.5514,3.1062] | [1.7062,3.3719] | [0.0000,0.0000] | [0.7319,2.3019] | [1.4862,2.7729] | [1.4186,2.8486] |
| E1 | [1.5062,3.2814] | [1.7376,3.2257] | [1.3405,2.9086] | [1.0914,2.9724] | [0.0000,0.0000] | [1.2614,2.1400] | [1.8000,3.5300] |
| E2 | [1.6500,3.2100] | [1.4400,3.1700] | [1.7000,3.3100] | [1.1400,2.9700] | [0.8700,2.7100] | [0.0000,0.0000] | [1.0100,2.7900] |
| E3 | [1.6900,3.2300] | [1.6900,3.2500] | [1.4400,3.0700] | [1.3800,3.0000] | [1.8000,3.1600] | [1.3500,2.9700] | [0,0000,0.0000] |

**Table A3.** The normalised rough group direct-relation matrix

| Barriers | I1 | I2 | I3 | I4 | E1 | E2 | E3 |
|---|---|---|---|---|---|---|---|
| I1 | [0.0000,0.0000] | [0.0643,0.1153] | [0.0612,0.1135] | [0.0475,0.1087] | [0.0596,0.1178] | [0.0528,0.1149] | [0,0428,0.1015] |
| I2 | [0.0638,0.1250] | [0.0000,0.0000] | [0.0503,0.1059] | [0.0472,0.1016] | [0.0445,0,1020] | [0.0536,0.1127] | [0,0546,0.1200] |
| I3 | [0.0633,0.1101] | [0.0600,0.1146] | [0.0000,0.0000] | [0.0584,0.1121] | [0.0340,0.0864] | [0.0638,0.1089] | [0,0471,0.1082] |
| I4 | [0.0638,0.1250] | [0.0549,0.1099] | [0.0604,0.1193] | [0.0000,0.0000] | [0.0259,0.0814] | [0.0526,0.0981] | [0,0502,0.1008] |
| E1 | [0.0533,0.1161] | [0.0615,0.1141] | [0.0474,0.1029] | [0.0386,0.1052] | [0.0000,0.0000] | [0.0446,0.0757] | [0,0637,0.1249] |
| E2 | [0.0584,0.1136] | [0.0510,0.1122] | [0.0602,0.1171] | [0.0403,0.1051] | [0.0308,0.0959] | [0.0000,0.0000] | [0,0357,0.0987] |
| E3 | [0.0598,0.1143] | [0.0598,0.1150] | [0.0510,0.1086] | [0.0488,0.1061] | [0.0637,0.1118] | [0.0478,0.1051] | [0,0000,0.0000] |

**Table A4.** The rough total-relation matrix

| Barriers | I1 | I2 | I3 | I4 | E1 | E2 | E3 |
|---|---|---|---|---|---|---|---|
| | Lower | Lower | Lower | Lower | Lower | Lower | Lower |
| I1 | 0,0272 | 0,0870 | 0,0826 | 0,0667 | 0,0759 | 0,0741 | 0,0634 |
| I2 | 0,0861 | 0,0254 | 0,0719 | 0,0656 | 0,0618 | 0,0738 | 0,0730 |
| I3 | 0,0867 | 0,0828 | 0,0250 | 0,0766 | 0,0525 | 0,0840 | 0,0667 |
| I4 | 0,0858 | 0,0769 | 0,0807 | 0,0204 | 0,0443 | 0,0727 | 0,0683 |
| E1 | 0,0763 | 0,0830 | 0,0687 | 0,0575 | 0,0192 | 0,0653 | 0,0812 |
| E2 | 0,0784 | 0,0709 | 0,0781 | 0,0572 | 0,0468 | 0,0206 | 0,0532 |
| E3 | 0,0838 | 0,0832 | 0,0735 | 0,0680 | 0,0802 | 0,0696 | 0,0226 |

**Table A5.** Row and column sums of the rough total-relation matrix

| | $x_i$ | | $y_j$ | |
|---|---|---|---|---|
| | Lower | Upper | Lower | Upper |
| I1 | 0.5243 | 2.0035 | 0.4769 | 1.9200 |
| I2 | 0.5092 | 1.9446 | 0.4576 | 1.9100 |
| I3 | 0.4806 | 1.9093 | 0.4743 | 1.8397 |
| I4 | 0.4120 | 1.8356 | 0.4490 | 1.8255 |
| E1 | 0.3808 | 1.7241 | 0.4512 | 1.8374 |
| E2 | 0.4599 | 1.7803 | 0.4052 | 1.8452 |
| E3 | 0.4284 | 1.8735 | 0.4809 | 1.8930 |